\documentclass[conference]{IEEEtran}

\ifCLASSINFOpdf
  \usepackage{graphicx}
  \usepackage[colorlinks=true,linkcolor=blue,citecolor=blue,pdfmenubar=true]{hyperref}
  \usepackage{rotating}
  \usepackage{booktabs}
  \usepackage{multirow}

\else

\fi

\begin{document}

\title{Design of a Mobile Face Recognition System for Visually Impaired Persons}

\author{
\IEEEauthorblockN{Shonal Chaudhry\IEEEauthorrefmark{1}\IEEEauthorrefmark{2} and Rohitash Chandra\IEEEauthorrefmark{1}\IEEEauthorrefmark{2}}
\IEEEauthorblockA{\IEEEauthorrefmark{1}School of Computing, Information and Mathematical Sciences, \\ Faculty of Science, Technology and Environment, \\ 
    The University of the South Pacific, Suva, Fiji}
\IEEEauthorblockA{\IEEEauthorrefmark{2}Artificial Intelligence and Cybernetics Research Group, \\ Software Foundation, Nausori, Fiji}
}

\maketitle

\begin{abstract}
It is estimated that 285 million people globally are visually impaired. A majority of these people live in developing countries and are among the elderly population. One of the most difficult tasks faced by the visually impaired is identification of people. While naturally, voice recognition is a common method of identification, it is an intuitive and difficult process. The rise of computation capability of mobile devices gives motivation to develop applications that can assist visually impaired persons. With the availability of mobile devices, these people can be assisted by an additional method of identification through intelligent software based on computer vision techniques. In this paper, we present the design and implementation of a face detection and recognition system for the visually impaired through the use of mobile computing. This mobile system is assisted by a server-based support system. The system was tested on a custom video database. Experiment results show high face detection accuracy and promising face recognition accuracy in suitable conditions. The challenges of the system lie in better recognition techniques for difficult situations in terms of lighting and weather.
\end{abstract}

\section{Introduction}
It is estimated that 285 million people globally are visually impaired\footnote{This is a general term for people with vision based disabilities. People who are blind have no vision while those with low vision have limited sight.} with 39 million blind and 246 million with low vision \cite{WHO_Webpage}. Approximately 90{\%} of these people live in developing countries and 82{\%} of blind people are aged 50 and above. Visually impaired persons adapt to life by using various assistive methods such as the white cane, sensory substitution and electronic devices \cite{Boerner_Social_support_and_well_being, Wan_Early_but_not_late_blindness}. The white cane is a common mobility tool used by the visually impaired \cite{White_Cane}. Sensory substitution is accomplished by using one sense to compensate for the lack of another. Research has shown that individuals which are blind from an early age have enhanced hearing compared to those with late blindness and to sighted individuals due to early initiation of sensory substitution \cite{Wan_Early_but_not_late_blindness, Gougoux_Voice_perception_in_blind}.

Electronic assistive systems use sensors or other methods to aid visually impaired users \cite{Willis_RFID_Information_Grid, Bahadir_Wearable_obstacle_detection_system}. They normally concentrate on providing navigation in indoor, outdoor or both environments \cite{Stewart_Accessible_Contextual_Information, Fernandes_Location_based_Services}. Navigation is commonly provided through the use of systems which employ some form of computer vision to detect obstacles, paths and perform location determination. These systems often utilize Global Positioning System (GPS) devices and web-based location services \cite{Fernandes_Location_based_Services} when providing navigation in outdoor environments due to their high accuracy \cite{Stewart_Accessible_Contextual_Information}. For indoor navigation, various types of Radio-Frequency Identification (RFID) based systems have been used to assist users in navigation by supplementing computer vision systems \cite{Fernandes_Location_based_Services, Falco_Indoor_Navigation_Multi_agent_System, Dudas_ONALIN}. Most RFID based systems use passive tags which are inexpensive and do not require a power source when not in use \cite{Chawla_An_overview_of_passive_RFID}.

While there are several systems for assisting the visually impaired in navigation, there are relatively few systems which help them locate and identify specific objects \cite{Willis_RFID_Information_Grid, Saleiro_SmartVision, Praveen_Blind_Navigation_Assistance}. Object detection and recognition are among the most difficult tasks faced by visually impaired persons. Object detection is the process of locating objects in a given environment \cite{Jain_Machine_Vision_1995}.

An object detection system was proposed by Al-Khalifa in \cite{Khalifa_Utilizing_QR_Code}. This barcode based system assisted users in identifying objects at museums and shopping centres. The system used a mobile phone camera to scan Quick Response (QR) codes attached to an object. Once the barcode was decoded, the phone's browser would retrieve an audio file containing a description of the object from the internet and playback its content to inform the user. Tian et al. presented another object detection system which supported users by detecting the presence of doors \cite{Tian_Computer_Vision_based_Door}. The system consisted of a camera and a computer. Features such as edges and corners were utilized to detect doors and distinguish them from other similar objects. In addition, the proposed method was found to be robust against variations in colour, texture, obstructions, illumination, scale and viewpoints. Experiments done by the authors showed that the algorithm achieved an accuracy of 91.9{\%} with a false positive detection rate of 2.9{\%}.

The object recognition process expands on object detection by identifying the detected object \cite{Jain_Machine_Vision_1995, Liter_An_introduction_to_object_recognition}. To accomplish this, object recognition systems are trained on a training dataset that determines which objects will be recognised. This allows object recognition systems to be used in a variety of applications such as counting, sorting and scene categorization \cite{Treiber_An_Introduction_to_OR}. Some of the challenges faced by object recognition systems are variances in lighting, rotation, shape and size of objects \cite{Palaniv_Multimodal_person_authentication}.

Nanni and Lumini developed an object recognition system in \cite{Nanni_Heterogeneous_bag_of_features}. The system was an improvement on a previous object recognition approach by the same authors \cite{Nanni_Random_Interest_Regions_for_Object}. To achieve improvements over the previous approach, the authors used local and global descriptors to represent images, performed dimensionality reduction of the texture descriptors using PCA-SIFT \cite{Ke_PCA_SIFT} and using a bag-of-words (BoW) approach to compute textons among other methods. According to the authors, textons refer to fundamental micro-structures in generic natural images and the basic elements in early (pre-attentive) visual perception. Overall, the new approach was able to achieve a high accuracy when applied to object, scene and landmark datasets.

Visually impaired persons can be assisted in object detection and recognition through the use of electronic devices. These devices must be small enough to be easily portable and have the necessary hardware capable of performing object detection and recognition \cite{Jafri_Face_recognition_for_the}. Mobile devices in particular are able to provide both these features and have already been used to deliver health services to users \cite{Bayu_Nutritional_Information_Visualization,Kranz_The_mobile_fitness_coach,Fairhurst_Texting_appointment_reminders}. Applications in the area of health have utilized hardware and software approaches such as cameras, sensors (accelerometers, gyroscopes, etc.) and text messaging to improve health care of users \cite{Klasnja_Healthcare_in_the_pocket}.

Cameras have become standard in almost every mobile device sold today \cite{Bourke_The_Social_Camera}. They have been used to provide nutritional information through augmented reality or recommend food intake \cite{Bayu_Nutritional_Information_Visualization, Kong_DietCam}. Modern mobile devices also have at least one type of built-in sensor. The device's sensors are commonly used for a type of health application known as fitness applications \cite{Buttussi_Bringing_Mobile_Guides_and_Fitness}. Fitness applications \cite{Kranz_The_mobile_fitness_coach, Consolvo_Design_Requirements_for_Technologies} can provide information on the amount of physical activity performed, give exercise feedback and motivate the user to keep a regular exercise schedule. Text messaging is a basic service available on all types of mobile devices with SIM (subscriber identity module) card support and can be used to deliver simple and effective health services to users \cite{Waller_Participatory_design_of_a_text_message}. Health systems can use text messaging to send appointment reminders to users or provide a support system to patients \cite{Fairhurst_Texting_appointment_reminders, Franklin_A_randomized_controlled_trial}.

The use of mobile applications does not necessarily need to be limited to sighted people. When object recognition is applied to faces, it can be used for identification. Face recognition is a suitable method for assisting visually impaired persons in identifying people compared to other biometric methods such as fingerprint, iris and voice recognition \cite{Jain_An_introduction_to_biometric, Kang_Mobile_iris_recognition, Jafri_A_Survey_of_Face_Recognition}. The advantages of face recognition over other approaches are passive recognition ability, fewer data acquisition errors and low implementation costs. Face recognition has been successfully used in previous research to create portable assistive systems for the visually impaired \cite{Jafri_Face_recognition_for_the,Sreekar_A_wearable_face_recognition,Kramer_Smartphone_based_face_recognition}. Other areas where face recognition has been applied include electoral registration, identification cards, security, social networking and surveillance \cite{Jafri_A_Survey_of_Face_Recognition, Ortiz_Face_recognition_for_web}.

With the rapid increase in computation power of mobile devices over the last decade, complex face recognition systems can be made portable. This paper proposes development of a face detection and recognition system using a mobile device. The system captures images of a person's face and matches it to faces in an existing database. Once a person is identified, the user of the mobile application is informed. The paper gives a framework of the overall system design and then gives a simple prototype implementation using built-in algorithms from the OpenCV (Open Source Computer Vision) library \cite{OpenCV_Home}. This prototype system is then evaluated by conducting various tests. It also demonstrates the integration of OpenCV\footnote{OpenCV is a computer vision and machine learning software library. It was built to provide a common infrastructure for computer vision applications and to increase the use of machine perception in commercial products \cite{OpenCV_About}.} with the Android \cite{Android_Home} mobile operating system. The proposed system is designed to work on different mobile devices that include tablets and smartphones. 

The rest of the paper is organised as follows. Background information is reviewed in Section 2 of the paper while Section 3 discusses the system's design. Section 4 provides development results, testing results and implementation details of the system. Implementation issues and limitations of the system are discussed in Section 5. Finally, the paper is concluded in Section 6 with a description of future work.

\section{Background}
\subsection{Assistive Systems}
There are many assistive systems in existence which are designed help visually impaired people. The most common type of system are those that provide navigation \cite{Praveen_Blind_Navigation_Assistance, Moreno_Realtime_Local_Navigation}. Willis and Helal presented a navigation and location determination system for the blind using an RFID tag grid \cite{Willis_RFID_Information_Grid}. The system consisted of RFID tags programmed with coordinates and descriptions of the surroundings. A mobile device was used to perform computational operations while a RFID reader integrated into a walking cane and shoe read RFID tags. This configuration had the advantage of not having to rely on a database or wireless infrastructure for access to information.

In \cite{Praveen_Blind_Navigation_Assistance}, Praveen and Paily proposed a different approach to navigation. They used a depth estimation technique from a single image based on local depth hypothesis. Their approach used a camera to capture an image of the environment in front of the user. After the image was captured, obstacles were then isolated using edge detection and morphological operations. Next, the depth was estimated for each obstacle using local depth hypothesis. Afterwards, the estimated depth map was compared with the reference depth map of the corresponding depth hypothesis. The difference between the estimated and reference depth map was then used to retrieve spatial information about the obstacles ahead of the user.

Moreover, a contextual information system for hearing and vision impaired people was presented by Stewart et al. in \cite{Stewart_Accessible_Contextual_Information}. The system, known as \textit{Talking Points}, provided contextual information about points of interest (POI) along a person's route. It consisted of a mobile device for accessing information, an online database for storing POI information, POI tags (Bluetooth beacons) and software with graphical and speech interfaces. A mobile device allowed easy access to information from the online database and precise detection of POI tags through the built-in Bluetooth interface. Addition of new content and updates to the online database were done by community contributors.

In addition to navigation systems, there are also assistive systems which detect objects. A wearable obstacle detection system was proposed by Bahadir et al. in \cite{Bahadir_Wearable_obstacle_detection_system}. This system was fully integrated into clothing for ease of use and consisted of ultrasonic sensors, vibration motors, power supplies and a micro-controller. The two main features of the system were detection of obstacles using sensors and guiding the user through an algorithm based on a neuro-fuzzy controller \cite{Jang_Neuro_fuzzy_modeling}.

Morelli and Folmer developed another system that enabled vision impaired users to play video games \cite{Morelli_Real_time_sensory_substitution}. The system used real-time video analysis to detect visual cues in a gesture-based video game and provided users with vibrotactile cues instead. Video feed from the game was sent to a laptop computer which utilized Extensible Markup Language (XML) based configuration files for determining the location of visual cues. Studies carried out by the researchers showed no major difference in gaming capability between sighted and vision impaired players.

Some implementations combine navigation and detection of objects to create advanced assistive systems. An example of such a system is the \textit{SmartVision} prototype by Hans du Buf et al. \cite{Saleiro_SmartVision}. This low cost navigation aid was designed to complement the use of white canes through the use of mobile devices. The system detected objects, obstacles and paths through a combination of computer vision, Geographic Information System (GIS), GPS and Wi-Fi. These modules were used to track the user's current location, plan routes and provide information about nearby POI. This system was further enhanced by the \textit{Nav4B} prototype which used RFID tags to address the limitations of the \textit{SmartVision} system \cite{Fernandes_Location_based_Services}.

A successor to the \textit{SmartVision} navigation aid was developed by Moreno et al. in \cite{Moreno_Realtime_Local_Navigation} as part of the \textit{Blavigator} project. The new system added detection of doors in corridors and a sound interface in addition to existing features. The sound interface assisted users by guiding them to the center of paths and alerting them to approaching obstacles. For outdoor navigation, the authors implemented two layer obstacle detection and avoidance \cite{Costa_Obstacle_Detection_using_Stereo}. The first layer was used for object detection while the second provided trajectory correction and backup. With the use of stereo vision, the authors implemented range image segmentation to extract information for object detection and recognition.

\subsection{Mobile Application Testing}
One of the key challenges of mobile application testing is performing tests on different devices since they differ in hardware configuration \cite{Amalfitano_Testing_Android_Mobile_Applications}. Some differences include processors (dual-core, quad-core, hexa-core or octa-core), memory (512 MB, 1 GB, etc.), storage space (4 GB, 8 GB, etc), input methods (touch screen or physical keypad), hardware sensors and screens with different display technologies.

The operating system of mobile devices also presents a challenge in testing because applications are handled differently by each operating system. For instance, an application may be easier to test on the Android operating system due to its open-source nature \cite{Ntantogian_Evaluating_the_privacy_of_Android}. Another difficulty faced in mobile application testing is availability of resources. Applications can behave differently when resources such as memory, battery or network connectivity are limited or unavailable.

According to Amalfitano et al., mobile applications are often tested by developers in three stages \cite{Amalfitano_Testing_Android_Mobile_Applications}. In the first stage, the application is tested on the development machine in an isolated environment using emulators. This allows the developers to quickly detect and correct implementation errors while the application is being developed. In the next stage, testing is done several times with some isolated features being removed in each iteration. After testing on emulators, the application is tested on a real device in the final stage to ensure compatibility.

Another approach to testing mobile application is the use of specialized testing software. These software simplify the testing process by gathering information on specific components or automating specific tests. An example of a testing framework for mobile devices was proposed by Satoh \cite{Satoh_A_Testing_Framework}. The framework, known as \textit{Flying Emulator}, introduced a mobile agent-based emulator of a mobile device. It performed application-transparent emulation of a mobile application written in Java for a specific device. Since the emulator is a mobile agent, it can test the application in the environments of different networks. In \cite{Bo_MobileTest}, Bo et al. presented a tool for supporting automatic black-box testing of mobile applications. The tool known as \textit{MobileTest} used a sensitive-event based approach to simplify design of test cases, increase testing efficiency and reusablity. Experiments conducted by the authors on three different phone models showed the tool to be effective.

Furthermore, Hu and Neamtiu proposed a test automation approach \cite{Hu_Automating_GUI_Testing} for addressing Android-specific bugs with a focus on those that affect the GUI (graphical user interface). They collected and categorized bugs from 10 popular open-source Android applications. They observed that while bugs related to application logic were still present, the rest of the bugs were specific to the Android platform due to their general structure. The test approach created by the researchers combined automatic event and test case generation with runtime monitoring and log file analysis. Results showed the technique to be effective for activity, event and type errors. In \cite{Ridene_A_Model_driven_Approach}, an approach for automation of mobile application testing was presented by Ridene and Barbier. It consisted of a Domain-Specific Modelling Language (DSML) known as MATeL (Mobile Applications Testing Language), a pre-existing test bed and its software infrastructure. MATeL allowed description of test scenarios where similarities and differences between mobile phones could be clearly expressed. This allowed easy testing on different phone models.

\section{Proposed System Design}
\begin{figure*}[tb]
\centering
\includegraphics[scale = 0.475]{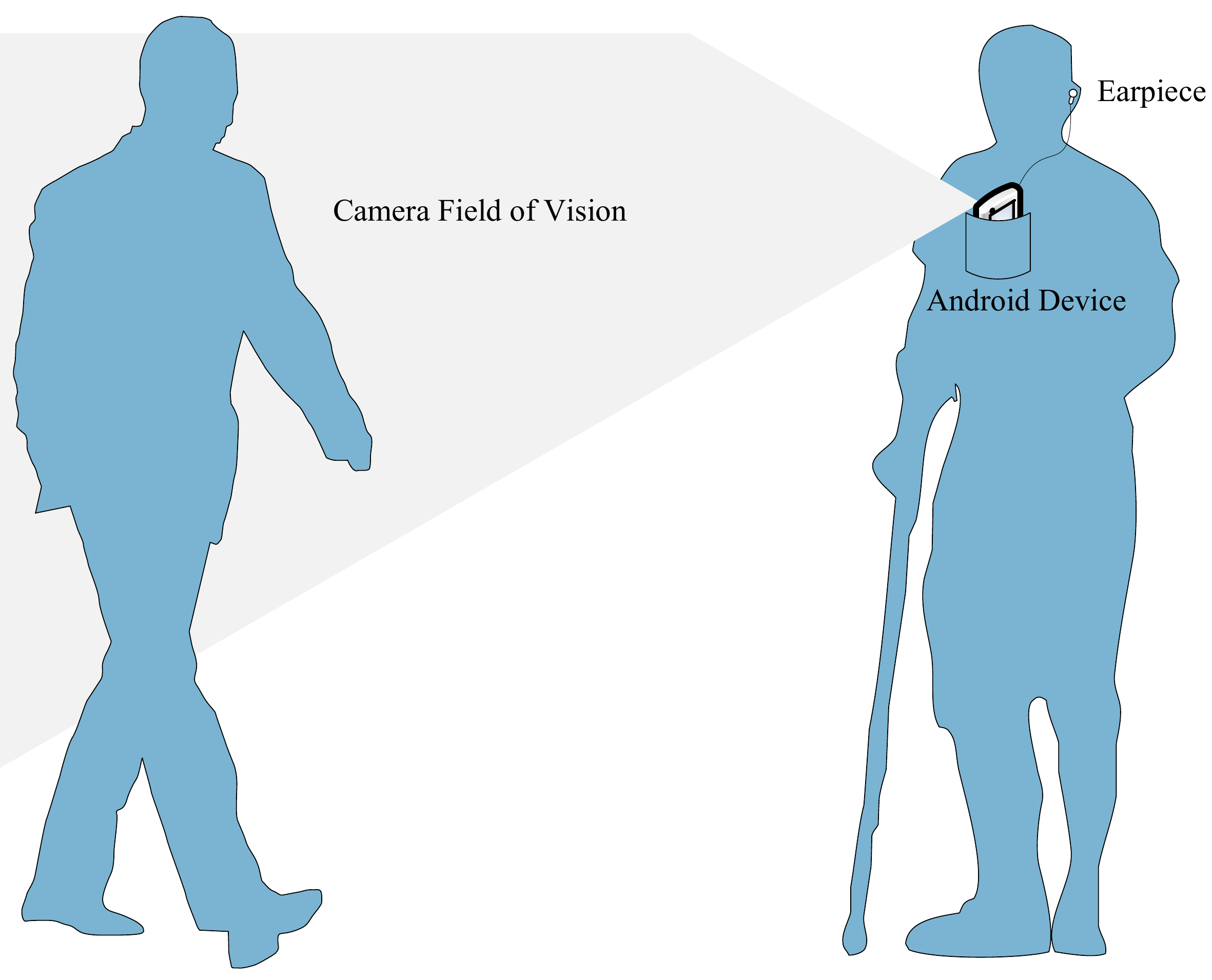}
\caption{Visualization of the proposed system. A camera is used to provide a wireless video feed to the mobile device.}
\label{Fig:Visualization}
\end{figure*}

Identification of people is a major challenge faced by the visually impaired. The increase in computation capability of mobile devices gives motivation to develop applications that can assist visually impaired persons. The proposed face recognition system is designed to take advantage of the portability of mobile devices and provide a simple user interface that makes use of the system easy for the visually impaired. Key design requirements for a portable system \cite{Saleiro_SmartVision, Lukianto_Stepping_Smartphone_based_portable} include small device size and low weight. To achieve this goal, a mobile device's camera and earpiece are used to form a compact and lightweight system (Fig \ref{Fig:Visualization}). In order to provide a convenient software experience for the portable system, a straightforward application in a preferably familiar operating system is needed. The Android operating system allows a user-friendly application to be developed through the use of built-in accessibility features.

\subsection{Software}
\subsubsection{Mobile Application}
The face detection and recognition application is intended to assist visually impaired users in locating and identifying people that they know. It features two states of operation; \textit{offline} and \textit{online}. These states correspond to the absence and presence of an internet connection respectively.

When in the \textit{offline} state, the application accesses the video feed from the device's camera. This video feed is then scanned to detect faces. When a face is detected, a temporary image is captured and saved on the device. After the image is saved, it is used to identify the person by searching for a possible match in the application's internal database. As soon as the person is identified, the result is displayed to the user. In the \textit{online} state, the video feed is still accessed for scanning and detecting faces. A temporary image is captured and saved when a face is detected in the same manner as in the \textit{offline} state. However, instead of attempting to identify the person, the image is sent to the support servers for identification. After the person is identified, the results are sent back to the application and displayed to the user.

Furthermore, the application allows a new person to be added to its internal face database via an option known as \textit{Enrolment Mode}. The internal database consists of a SQLite\footnote{https://www.sqlite.org/about.html} database and face images. Each face image is associated with one entry in the SQLite database. This internal database is also limited to the most frequently used faces for recognition. The number of faces to store in the database is defined by the user. In \textit{Enrolment Mode}, instead of performing identification after an image is captured, a new interface is displayed asking the user to enter details of the person. Upon entering the details of the person, their face is added to the database. The face detection and recognition features of the application are divided into two modules; \textit{Detection System} and \textit{Recognition System}.

\begin{figure}[hb]
\centering
\includegraphics[scale = 0.33]{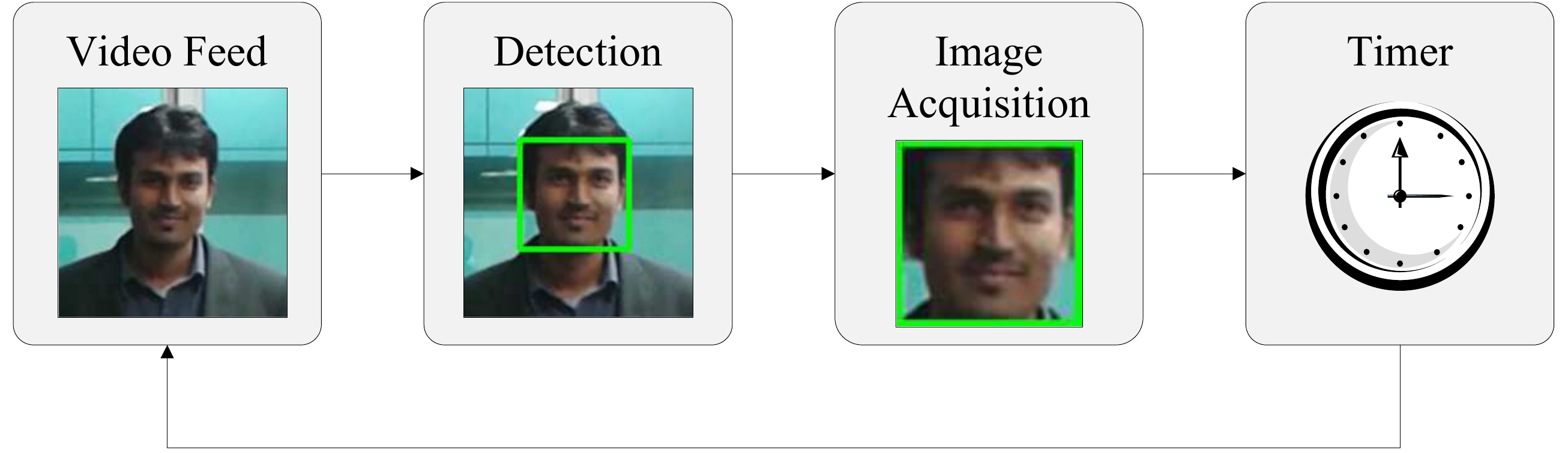}
\caption{Detection procedure of the mobile application.}
\label{Fig:Detection_System}
\end{figure}

\paragraph{Detection System}
The main goal of the detection system in the application is to detect faces. When a face is detected, a bounding box is drawn around it (Fig \ref{Fig:Detection_System}). This bounding box is used to extract and save the face of a person by cropping the area inside it. Once the face is detected, the next detection is performed after a delay to avoid overwhelming the user with constant detection notifications. Faces detected by the system are limited to frontal faces since the application is designed to identify persons in front of the user.

Detection is accomplished through the use of the object detector built into the OpenCV Android library \cite{OpenCV_Android}. The object detector uses a cascade classifier to detect faces. The cascade classifier is made up of a cascade of boosted classifiers with Haar-like features \cite{OpenCV_Cascade_Classifier}. Training of the classifier is done using a set of positive and negative images. In the case of face detection, the positive images are those which contain faces while the negative images are made of objects that are not faces. 

OpenCV's object detector is based on the work initially proposed by Viola \cite{Viola_Rapid_object_detection} and improved by Lienhart \cite{Lienhart_An_extended_set_of_haar}. The cascade in \emph{cascade classifier} refers to the resulting classifier created from many smaller classifiers (known as stages) which are applied sequentially to a specific region \cite{OpenCV_Cascade_Classifier}. \emph{Boosted classifiers} refer to classifiers at each stage of the cascade which are created from simple classifiers using a boosting technique \cite{Schapire_The_Strength_of_Weak}. Boosting is an algorithm for reducing bias in supervised learning. It works by combining weak classifiers to create a strong classifier.

\begin{figure}[tb]
\centering
\includegraphics[scale = 0.33]{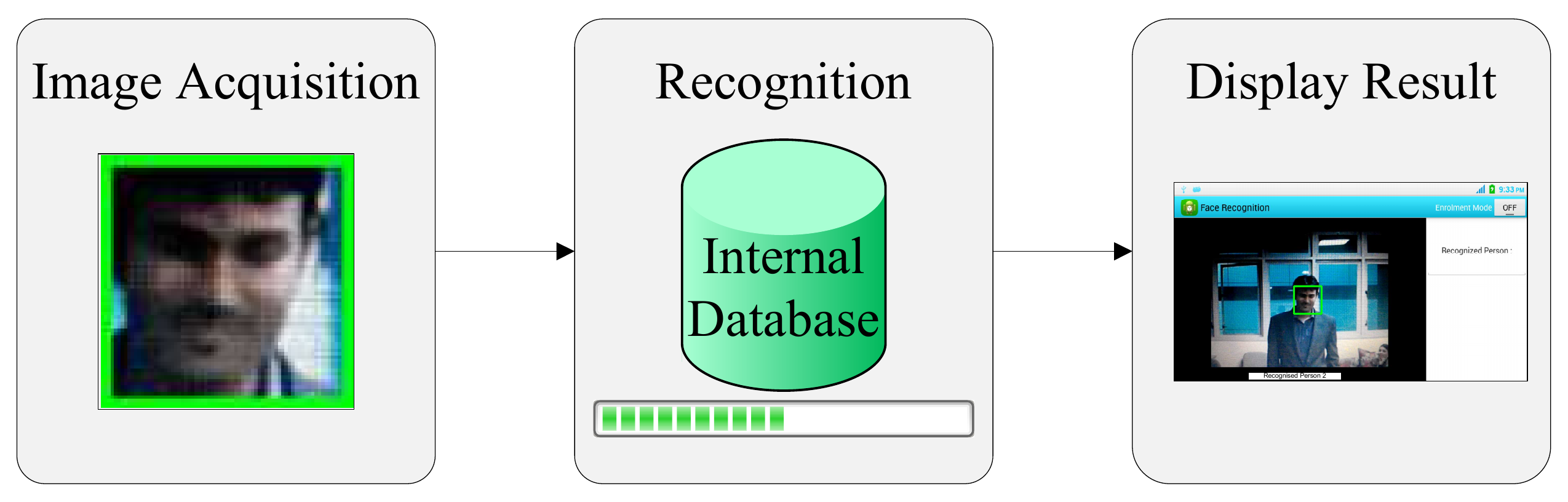}
\caption{Recognition procedure of the mobile application.}
\label{Fig:Recognition_System}
\end{figure}

\paragraph{Recognition System}
The recognition system (Fig \ref{Fig:Recognition_System}) complements the detection system by identifying the person using the detected face. This is done by running the recognition program which searches the internal database for a match using the saved face image as input. The recognition program utilizes C++ functions from the OpenCV library to identify people. This is due to limited support for Java-based face recognition in the OpenCV Android library. To enable use of the C++ functions, the Java Native Interface\footnote{http://docs.oracle.com/javase/7/docs/technotes/guides/jni/} (JNI) is used by the application.

The recognition program uses the Local Binary Patterns Histograms (LBPH) algorithm to perform recognition. It uses the Local Binary Patterns (LBP) method which creates a summary of the local structure in an image by comparing each pixel with its neighbours \cite{Ahonen_Face_recognition_with_local}. If the intensity of the centre pixel is greater or equal to its neighbour, it is assigned a value of 1 or 0 if not. The resulting binary numbers for each pixel are called Local Binary Patterns. In this approach, the LBP image is divided into local regions and a histogram is extracted from each image. The feature vector is then obtained by concatenating the local histograms which are called Local Binary Patterns Histograms. The use of the LBPH approach allows for fast feature extraction \cite{Ahonen_Face_recognition_with_local}.

\subsubsection{Support Servers}
The \textit{support servers} in the face recognition system refer to a server-based support system for the mobile application. They enable enrolment and identification of individuals through the mobile device's internet connection. This is accomplished by sending captured images and relevant information to the servers instead of processing it locally on the device. The implementation of support servers permits use of the application on low specification devices which do not have the hardware required for real-time face recognition. Lastly, the use of a server-based system can result in power savings on both low and high specification devices since computational load is reduced \cite{Miettinen_Energy_Efficiency_of_Mobile}.

\subsection{Hardware}
\begin{figure*}[tb]
\centering
\includegraphics[scale = 0.4]{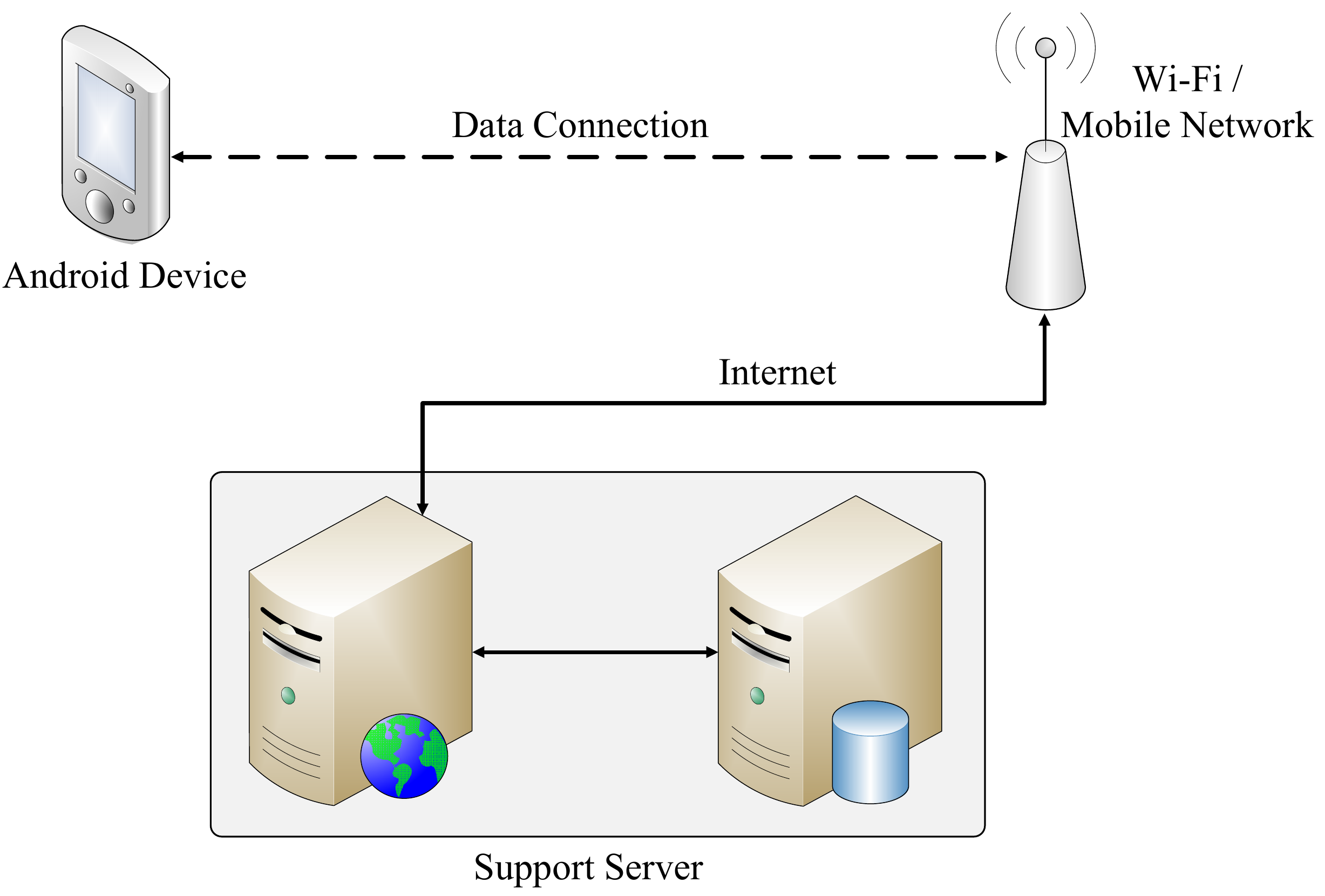}
\caption{Design of the mobile face recognition system.}
\label{Fig:System_Design}
\end{figure*}

The face recognition system consists of three hardware components; an Android device, internet infrastructure and support servers (Fig \ref{Fig:System_Design}). The Android device's camera is accessed by the application. For internet access, a mobile network data connection or Wi-Fi can be used. The configuration of the support servers depend on their implementation. Depending on the computation load, the number of servers can be increased as the user base grows.

In the \textit{offline} state, the application uses the video feed from the camera to detect and recognise faces. This is done in real-time using a device which can provide the necessary computation power. The prototype developed in this paper used an Android phone with a quad-core processor. While computation power requirements of the application can be satisfied by some mobile devices, the application will still be limited in operation time due to battery capacity.

The \textit{online} state conserves energy by reducing the computation load on the device. Energy efficiency can be increased by using context-aware power management techniques which alert the user if the device may run out of power before the next charging opportunity \cite{Ravi_Context_aware_Battery_Management}. Moreover, the type of display a device has also determines power consumption with some displays being more energy efficient than others \cite{Chen_How_is_Energy_Consumed_in_Smartphone}. For visually impaired users, the display on devices can be dimmed to the lowest level or turned off in some cases to further increase power savings since displays consume a significant amount of power \cite{Carroll_An_Analysis_of_Power_Consumption}.

\section{Results and Implementation}
The prototype system proposed in this paper shows how face recognition can be used to assist visually impaired persons. This section highlights the major components of the system, discusses the person identification process and presents results of experiments conducted on the system showing good accuracy. Furthermore, tests on the mobile device and its results are also discussed.

\begin{figure*}[ht]
\centering
\includegraphics[scale = 0.425]{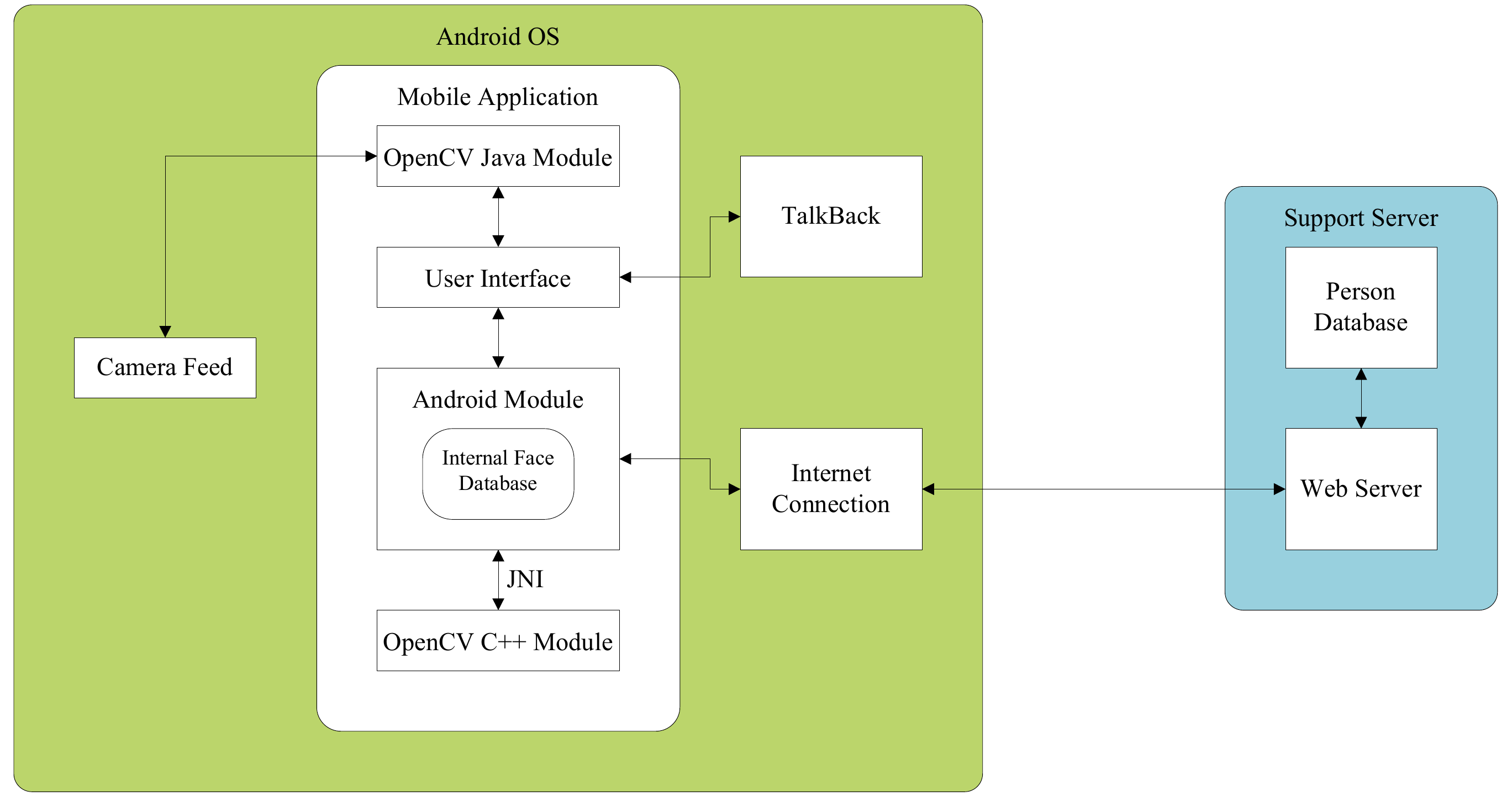}
\caption{Face recognition system software framework.}
\label{Fig:Software_Design}
\end{figure*}

The mobile application consists of four major components as shown in Fig \ref{Fig:Software_Design}. These are the OpenCV Java Module, User Interface, Android Module and OpenCV C++ Module. The OpenCV Java Module consists of the \textit{Detection System} and functions for interfacing with the Android Module. Relevant functions and layouts make up the User Interface while the Android Module consists of functions which manage the internal face database, interface with both OpenCV modules and handle overall operation of the application. Lastly, the OpenCV C++ Module contains of the \textit{Recognition System} and methods for interfacing with the Android Module.

The application interacts with three other components within the operating system to accomplish face recognition. These are the Camera Feed, Internet Connection and TalkBack \cite{Google_TalkBack}. \textit{Google TalkBack} is an accessibility service created by Google to assist visually impaired users with device interaction \cite{Google_TalkBack}. It adds spoken, audible and vibration feedback to a device. When the application is launched, the video feed from the device's camera is given to the OpenCV Java Module to perform face detection. After a face is detected, the OpenCV C++ Module is used to carry out face recognition. Once face recognition is complete, the results are conveyed to the user through the Android Module. Throughout the use of the application, TalkBack is used to provide audible information to the users allowing them to operate the application with limited sight.

\subsection{Prototype: Simulation and Experiments}
The mobile application was tested on a custom video database with a total of 8 videos. Each video was recorded using a mobile phone camera at a resolution of 1280 x 720 at 30 fps. All videos were taken at evening in artificial lighting conditions with variations in face orientations. The experiments were divided into two sections; \textit{Face Detection} and \textit{Face Recognition} with a total of 80 experiment runs for face detection and 50 experiment runs for face recognition respectively.

\subsubsection{Face Detection}
The results for face detection show detection accuracy of up to 93{\%} in well-lit conditions as shown in Table \ref{Table:Detection_Results}. Better detection accuracy was achieved when the person is looking directly at the camera with a neutral facial expression (Videos 1, 3 and 6). Lower accuracy is obtained when faces are detected at slight angles and different facial expressions are used (Videos 2, 4, 5 and 8). An ideal accuracy (Video 7) can be obtained when the person is walking directly toward the camera. This occurs when no false detections are made due to a simple background with no other objects. The result however is only possible when the user is stationary and ideal conditions are present. Computation times for each frame, whether a face was present or not, were less than 400 ms.

\begin{table}[htbp]
  \centering
  \caption{Results of face detection experiments.}
    \scalebox{0.95}{
    \begin{tabular}{c c c c c c}
    \toprule
    Video & Frames with Faces & Detections & Correct & Incorrect & Accuracy (\%) \\
    \midrule
    1     & 130   & 17    & 15    & 2     & 88.24 \\
    2     & 67    & 16    & 14    & 2     & 87.50 \\
    3     & 82    & 15    & 14    & 1     & 93.33 \\
    4     & 79    & 18    & 15    & 3     & 83.33 \\
    5     & 116   & 24    & 17    & 7     & 70.83 \\
    6     & 270   & 19    & 17    & 2     & 89.47 \\
    7     & 102   & 10    & 10    & 0     & 100.00 \\
    8     & 139   & 8     & 7     & 1     & 87.50 \\
    \bottomrule
    \end{tabular}
    }
  \label{Table:Detection_Results}
\end{table}

\subsubsection{Face Recognition}
Table \ref{Table:Recognition_Results} shows the results for face recognition which were obtained by feeding the results of face detection into the recognition program. However, only true detections (where an actual face was detected) were considered. Results for face recognition show relatively high recognition accuracy for Persons 1 and 4. This is possibly due to the person looking almost directly at the camera with a neutral face expression. Face recognition accuracy for Persons 2 and 3 are lower as a result of recognising faces at angles with different facial expressions. A major factor in low recognition accuracy is the use of only frontal face images for enrolling a person into the face database. These results indicate that further improvements to the face recognition program are required.

\begin{table}[htbp]
  \centering
  \caption{Results of face recognition experiments.}
    \begin{tabular}{c c c c c}
    \toprule
    Person & Experiments & Correct & Incorrect & Accuracy (\%) \\
    \midrule
    1     & \multirow{4}{*}{50} & 32    & 18    & 64.00 \\
    2     &       & 26    & 24    & 52.00 \\
    3     &       & 29    & 21    & 58.00 \\
    4     &       & 35    & 15    & 70.00 \\
    \bottomrule
    \end{tabular}
  \label{Table:Recognition_Results}
\end{table}

\subsection{Mobile Device Testing}
\subsubsection{Test Setup}
Battery usage is one of the most important factors in the performance of the face recognition application. Several components of the device are used when the application is in operation. Firstly, the application constantly receives frames from the video feed of the device's camera where energy is continually drained. Processing these frames to detect and recognise faces is also computationally expensive which further uses energy. Thirdly, energy is drained by the Wi-Fi internet connection when the support server is used to recognise faces. Lastly, feedback provided to the user on application navigation and recognition results continuously consumes energy. The battery use by the device is analysed by running the application and using Android's built-in battery application to obtain the power consumption. The power consumption from the battery application is verified using the Android Debug Bridge (ADB) \cite{Android_ADB}.

The impact on processing power by the face recognition application is analysed next. This is done by measuring the processor use through the ADB when the application is running (attempting to detect faces) and performing face recognition. Another factor which affects the performance of the application is memory use. The memory use is also measured by the ADB under the same scenarios as the processor use.

Analysing data use (internet usage) by the application is important since the use of mobile data can be expensive for the user. Data usage from performing server-based face recognition is measured using the built-in data monitoring application of the operating system and verified through the ADB.

\subsubsection{Testing Results}
\paragraph{Battery}
Battery usage by the device over time is shown in Table \ref{Table:Battery_Results}. Since the drain on energy by the application is significant due to real-time computer vision, the system is currently suitable for short-term use. In the experiment, energy use by the device was constant over time as the mobile application was running (Table \ref{Table:Battery_Percentage}). These values will change in real-world scenarios where the user may use other applications.

The mobile face recognition application accounts for 36{\%} of total battery used by the device. Energy used by the mobile network is 24{\%} of the total battery use while the Wi-Fi connection uses 17{\%}. The media server of the device uses the least energy at 3{\%}. Finally, a significant amount of energy (20{\%}) is used by the device when it is idle. This is due to some cores of the processor being idle while the application is running. These idle cores can allow the user to use other applications without needing to close the face recognition application for processing power.
\begin{table}[htbp]
  \centering
  \caption{Battery life of the device over time.}
    \begin{tabular}{c c}
    \toprule
    Time Elapsed & Battery Life (\%) \\
    \midrule
    10 mins & 95 \\
    20 mins & 92 \\
    30 mins & 90 \\
    45 mins & 84 \\
    1 hr 10 mins & 76 \\
    1 hr 25 mins & 71 \\
    1 hr 30 mins & 68 \\
    \bottomrule
    \end{tabular}
  \label{Table:Battery_Results}
\end{table}
\begin{table}[htbp]
  \centering
  \caption{Battery use of the device at any given time.}
    \begin{tabular}{c c c c c}
    \toprule
    Component & Battery Usage (\%) \\
    \midrule
    Application & 36 \\
    Mobile Standby & 24 \\
    Phone Idle & 20 \\
    Wi-Fi & 17 \\
    Media Server & 3 \\
    \midrule
    Total & 100 \\
    \bottomrule
    \end{tabular}
  \label{Table:Battery_Percentage}
\end{table}

\begin{figure*}[ht]
\centering
\includegraphics[scale = 0.5]{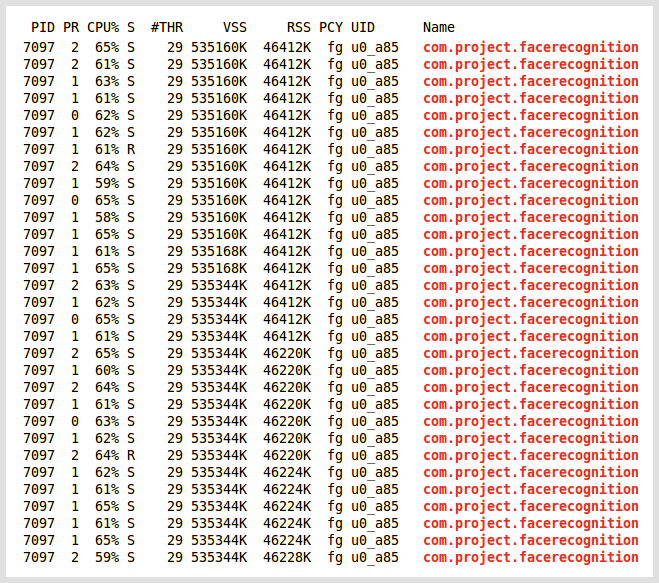}
\caption{Processing power used by the application.}
\label{Fig:Processor_Results}
\end{figure*}

\begin{figure*}[ht]
\centering
\includegraphics[scale = 0.55]{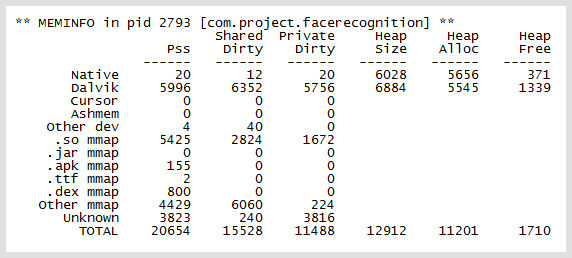}
\caption{Memory used by the application.}
\label{Fig:Memory_Results}
\end{figure*}

\paragraph{Processing Power \& Memory}
The processing power (CPU {\%}) used by the application is shown in Fig \ref{Fig:Processor_Results}. It shows 31 filtered entries taken by the ADB at 1 second intervals on a console. The minimum CPU use by the application is 59{\%}, the maximum is 65{\%} and the average use is 62{\%}.
Memory use by the application is shown in Fig \ref{Fig:Memory_Results}. The private memory use (i.e. used by the application only) is shown to be 11.488 MB. This indicates that the application is not memory intensive.

\begin{figure}[tb]
\centering
\includegraphics[scale = 0.3]{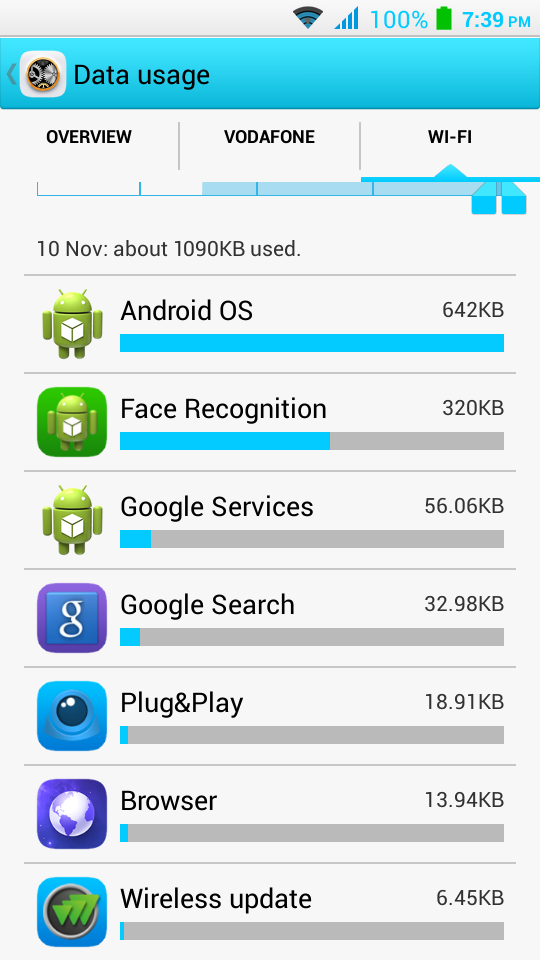}
\caption{Screenshot of data used by the application.}
\label{Fig:Data_Results}
\end{figure}

\paragraph{Data Usage}
The application does not use a large amount of data since most of the data sent and received consists of text and small images which are less than 100 kb. Data use when fully syncing a face database for one user account is shown in Fig \ref{Fig:Data_Results}. This full sync consists of 10 greyscale face images which have an average size of 20 kb and related text data on the person that each face represents.

\section{Discussion}
The mobile face recognition system was implemented on a 3G Android device with a 1.3 GHz quad-core processor and an 1800 mAh battery running Android 4.2.2. These specifications allow the device to process several video frames in real-time while maintaining application response. This device's camera provided a 1280 x 720 video feed. Major components of the mobile application included the OpenCV Java Module, User Interface, Android Module and OpenCV C++ Module. The face detection system was part of the OpenCV Java Module (to avoid increasing overhead from JNI calls when using C++ functions of OpenCV) while the recognition system was part of the OpenCV C++ Module. Finally, support servers assisted the application by providing fast enrolment and identification of people.

Deploying the mobile face recognition system for actual use will have challenges. A major concern is battery life of mobile devices since real-time computer vision constantly uses hardware resources. While the ability to offload computation tasks to the support servers can conserve power, this can lead to increased costs for the user due to the use of mobile data. In real-world deployment of the system, users may need to undergo training. Users may not be familiar with using an application while TalkBack is enabled. A tutorial is built into TalkBack for training users on how to use the application. The application does not require any maintenance as well since redundant files are automatically deleted. However, the \textit{support servers} may need an administrator to manage the face database when it grows to a very large size.

There may also be some difficulties with practical applications of the system. Security is a potential problem for the face recognition system. The most important area of concern in security is the face images of people and their identity. To ensure that these are protected, the face images are stored in a location accessible only to the application while the identities of people are stored in an encrypted format. The use of face recognition also creates a privacy issue since face images of people are used when performing recognition and enrolling individuals. To address this, the system captures a temporary image when performing recognition and stores face images in a folder only accessible by the application for enrolment. Once the recognition is complete or the internal database is updated with enrolled individuals, the face images are deleted in each case.

The mobile application was tested on a custom video database comprising of 8 videos with resolution of 1280 x 720 at 30 fps. A maximum face detection accuracy of 93{\%} and face recognition accuracy of 70{\%} were achieved in well-lit conditions. Since the goal of the system was to develop a simple prototype, comparisons with state-of-the-art algorithms and methods were not made. As a result, future work will be focused on improving the performance while keeping the computation requirements at a minimum.

In the current implementation, the face detection program is limited to only well-lit conditions. Experiments were conducted using poorly-lit versions of some videos from the custom database resulting in poor detection rates. The absence of bright light makes it difficult for the detector to correctly detect faces. Also, the test video database is small and limited to a few faces.

A future extension to the mobile application can assist sighted persons as well through the use of a zooming feature. A zooming feature can be used as a substitute for reading glasses by enlarging written text through the device's camera and displaying them to the user. Furthermore, the face detection program can be further refined to work in poorly-lit conditions. The test video database could also be expanded to include more faces under different lighting and weather conditions.

\section{Conclusion}
This paper discussed the development of a mobile based face recognition system. A mobile device's camera was used for image acquisition while an Android application processed the image. The application featured a face detection and recognition program.

The face detection program detected faces from the video feed using a cascade classifier. Once a face was detected, it was recognised using a temporary image of the  detected face. \textit{Google TalkBack} was used to provide users with feedback on the operation of the application. Experiments showed a face detection accuracy of 93{\%} and face recognition accuracy of 70{\%} in well-lit conditions. The performance of the mobile application was also evaluated by conducting tests on the mobile device. Results show low memory and data usage as strengths of the system while limitations include energy use and processing power.

Future work will focus on improving the detection and recognition systems. The detection system can be further improved by training a new cascade classifier for the face detector while a neural network based program can be used to improve the recognition system. In addition, the system can be tested with actual users to determine the positive and negative aspects of the system's design. Wearable devices such as smartwatches could also improve the system by making it easier for the user to interact with the mobile application.

\bibliographystyle{IEEEtran}
\bibliography{IEEEabrv,Mobile_Face_Recognition}

\end{document}